\documentclass[aps,pra,twocolumn]{revtex4-1}

\usepackage{graphicx}

\begin{document}

\title{Weak valued statistics as fundamental explanation of quantum physics
}

\author{Holger F. Hofmann}
\email{hofmann@hiroshima-u.ac.jp}
\affiliation{
Graduate School of Advanced Sciences of Matter, Hiroshima University,
Kagamiyama 1-3-1, Higashi Hiroshima 739-8530, Japan}
\affiliation{JST, CREST, Sanbancho 5, Chiyoda-ku, Tokyo 102-0075, Japan
}

\begin{abstract}
Recently, weak measurements have attracted a lot of interest as an experimental method for the investigation of non-classical correlations between observables that cannot be measured jointly.  Here, I explain how the complex valued statistics observed in weak measurements relate to the operator algebra of the conventional Hilbert space formalism and show that the algebra of operators originates from more fundamental relations between the physical properties of a quantum system that can be expressed in terms of complex conditional probabilities. In particular, commutation relations can be identified with fundamental imaginary correlations that characterize the relations between physical properties in terms of their transformation dynamics. Non-commutativity thus originates from a definition of relations between physical properties that replaces the assumption of joint reality with a complex-valued probability reflecting the dynamical response of the system to external forces, e.g. in measurement interactions.
\end{abstract}

\maketitle

\section{Introduction}

In the most general formulation of quantum mechanics, classical variables are replaced by Hilbert space operators acting on state vectors. If one of these observables is known, the system is considered to be in an eigenstate of the operator, which can be written as a superposition of any other set of eigenvectors. This is the point where the physical meaning is unclear. In classical physics, a system is described by simultaneously assigning values to all its properties. One might expect that such an assignment should also be possible in quantum mechanics, since an appropriate measurement will always result in a single precise outcome. Why is it not possible to replace the probabilities of measurement outcomes predicted by a quantum state by the actual values of the physical properties observed in the measurement? Clearly, the fact that operators cannot just be replaced by their eigenvalues indicates a very fundamental difference between the physics described by operators and the physics of classical variables. A proper explanation of quantum physics should identify this difference by explaining in detail how the various properties of a quantum system are related to each other, and how the results obtained in different measurements can be explained in terms of these fundamental relations. Unfortunately, the original formulation of operator algebra was not based on a general theory of measurements, but established the mathematical structure in terms of approximate analogies such as the wave-particle dualism, which are not sufficiently general to address all of the questions raised by a thorough investigation of quantum measurements. The consequence has been a string of seemingly paradoxical results that can be predicted and described using the standard theory, but appear to have no satisfactory explanation in terms of the underlying physics. 

It may seem strange that predictions of a widely accepted formalism have no physical explanation. However, quantum theory was constructed from very little empirical evidence, and it was not initially clear which details of the theory would ever be accessible to experimental tests. Based on this lack of evidence, the terminology was developed by analogies with classical concepts, even where these concepts directly contradicted the relations between their quantum mechanical analogs. This is most clearly seen in the operator formalism, e.g. when the spin is defined as a three dimensional vector, even though it is never possible to identify a direction based on a simultaneous assignment of three eigenvalues to the three non-commuting components. Curiously, the assignment of non-commuting operators to the three orthogonal components does reproduce the transformations associated with spatial rotations, and this seems to correspond to the relations between averages observed in measurements along different spatial directions. It is therefore possible to reduce the problem to a more specific question: how are the individual measurement outcomes represented by eigenstates related to the algebra of operators that reproduces the transformations between different ``orientations,'' or contexts, of a quantum measurement? 

If one recognizes that this is in fact an open question, it is possible to get new insights from the results recently reported in the context of weak measurements. In particular, weak measurements can directly establish a relation between three non-commuting properties of a quantum system by combining preparation and a precise final measurement with the intermediate measurement of a third property. Quantum states can then be characterized by complex joint probabilities of only two non-commuting properties, as long as the eigenstates of the two properties have non-zero mutual overlap \cite{Joh07,Lun11,Lun12,Hof11b,Hof12a}. Weak measurements thus provide a better understanding of quantum states in terms of fundamental statistics. However, the essential point might actually be that weak measurements can actually reveal fundamental laws of physics that do not depend on the specific quantum state and are universally valid in any context. As pointed out in \cite{Hof11b,Hof12a}, the complex conditional probabilities observed in weak measurements describe the fundamental relation of three physical properties in terms of transformations between eigenstates of two properties generated by the third. In the following, I intend to argue that these relations are the natural explanation of the operator algebra, and that the conventional operator formalism is best understood as a theory of complex conditional probabilities. 

\section{Operator physics and uncertainties}

The key observation that makes new insights into the quantum formalism possible is that the original Hilbert space formalism is more complete and more precise than the textbook examples suggest. In particular, it is possible to analyze the precision of statements about unobserved properties, as explained by Ozawa in his groundbreaking work on measurement uncertainties \cite{Oza03}. Specifically, Ozawa introduced a mathematical expression for the error of a measurement when neither the initial state nor the measurement results are represented by an eigenstate of the target observable. The very fact that such an expression exists proves that there is more to the operator formalism than the textbooks would tell. Interestingly, it was pointed out soon after by Hall that Ozawa's uncertainties are minimized by the real parts of the weak values \cite{Hal04}. Indeed, the proper conclusion seems to be that the results of weak measurements represent the zero uncertainty values of an operator for a specific selection of initial pure state and precise final measurement \cite{Hos10,Lun10,Hof12b}. Specifically the uncertainty of $\hat{A}$ in an initial state $\mid \psi \rangle$ can be fully explained by the fluctuations of the complex weak values observed in {\it any} final measurement with outcomes $\mid m \rangle$: 
\begin{equation}
\label{eq:fluct}
\langle \psi \mid \hat{A}^2 \mid \psi \rangle = \sum_m \left|\frac{\langle m \mid \hat{A} \mid \psi \rangle}{\langle m \mid \psi \rangle}\right|^2 |\langle m \mid \psi \rangle|^2.
\end{equation}
Since the weak values conditioned by $\mid \psi \rangle$ and $\mid m \rangle$ appear to provide a complete description of the quantum fluctuations of $\hat{A}$ in $\mid \psi \rangle$, it is possible to think of the operator $\hat{A}$ as a ``function'' determined by the initial state and the final state, even though the values of this function do not correspond to the eigenvalues of $\hat{A}$. In fact, this functional dependence of the value of $\hat{A}$ on a pair of non-commuting eigenstates was already introduced by Dirac as early as 1945 \cite{Dir45}, who thus discovered weak values 43 years before the actual introduction of weak measurements. However, there is a problem with this representation of an operator $\hat{A}$ as a function of two precise measurement outcomes. If the weak value were the true value of $\hat{A}$ for a reality jointly defined by $\mid \psi \rangle$ and $\mid m \rangle$, we would expect to find that the weak value of $\hat{A}^2$ would be equal to the square of the weak value. The operator formalism shows that this is not the case. Instead, weak values seem to have complex-valued uncertainties that average to zero in Eq. (\ref{eq:fluct}). 

The failure of a naive realist interpretation of weak values suggests a return to eigenvalues as an alternative. Using the spectral decomposition of the operator $\hat{A}$, the weak value can be written as an average defined by complex conditional probabilities,
\begin{equation}
\frac{\langle m \mid \hat{A} \mid \psi \rangle}{\langle m \mid \psi \rangle} = \sum_a A_a \; p(a|\psi,m),
\end{equation}
where $p(a|\psi,m)$ is given by the complex weak value of the projector $\mid a \rangle\langle a \mid$. Therefore, weak values are consistent with the eigenvalues $A_a$ if they are interpreted as an average of complex conditional probabilities that describe the relation between the measurement results $\psi$, $m$, and $a$. Significantly, it is not necessary to interpret these weak values as specific measurement results. Instead, they are elements of a measurement independent representation of $\hat{A}$ \cite{Hof12a}. It is therefore possible to express all mathematical relations between operators in terms of complex conditional probabilities, independent of the specific measurement context. 

Since the non-commutativity of operators is often emphasized as a key difference between quantum physics and classical physics, it may be worth taking a closer look at the complex statistics expressed by commutation relations. In particular, the uncertainty limit of two non-commuting observables for pure states $\mid \psi \rangle$ is often explained in terms of the expectation value of the commutation relation,
\begin{equation}
\label{eq:limit}
\Delta A \Delta B \geq \frac{1}{2} |\langle \psi \mid [ \hat{A}, \hat{B} ] \mid \psi \rangle|.
\end{equation}
Using the spectral decomposition of the operators $\hat{A}$ and $\hat{B}$, the commutation relation can be written as a sum over all possible combinations of measurement outcomes $a$ and $b$, with a complex joint probability of $\rho(a,b|\psi)$ obtained for the quantum state $\mid \psi \rangle$. The expectation value of the commutation relation then corresponds to the imaginary part of the complex correlation between $\hat{A}$ and $\hat{B}$ in $\rho(a,b|\psi)$, and the uncertainty limit of Eq.(\ref{eq:limit}) is given by
\begin{equation}
\label{eq:comrel}
\frac{i}{2} \langle \psi \mid  [ \hat{A}, \hat{B} ] \mid \psi \rangle 
=
\sum_{a,b} A_a B_b \;\; \mbox{Im}\left( \rho(a,b|\psi) \right).
\end{equation}
Non-commutativity can therefore be explained in terms of complex joint probabilities. The standard uncertainty limit given by Eq.({\ref{eq:limit}}) is then a natural consequence of the classical rule that correlations cannot be larger than the product of the fluctuations, which applies equally to complex and to real valued probabilities. 

It is possible to measure the complete complex probability $\rho(a,b|\psi)$, e.g. by performing a weak measurement of the projector on $\mid a \rangle$ followed by a final measurement of $b$ \cite{Lun12}. The joint probability is then given by the product of weak value and measurement probability,
\begin{eqnarray}
\rho(a,b|\psi) &=& p(a|\psi,b) |\langle b \mid \psi \rangle|^2
\nonumber \\
&=& \langle b \mid a \rangle\langle a \mid \psi \rangle \langle \psi \mid b \rangle.
\end{eqnarray}
Importantly, this joint probability expresses a highly symmetric relation between the three properties represented by $a$, $b$, and $\psi$. If we replace $\mid \psi \rangle$ with the eigenstates $\mid m \rangle$ of a third property $\hat{M}$, the relations between $\hat{A}$, $\hat{B}$ and $\hat{M}$ are determined by a set of $d^3$ joint probabilities that represent any of the three eigenstates in terms of joint probabilities of the two other eigenstates,
\begin{equation}
\rho(a,b|m) = \rho(m,a|b)=\rho(b,m|a).
\end{equation} 
Therefore, the initial state in the complex joint probability for a system with know property $m$ is fully determined by a fundamental relation between $m$, $a$ and $b$, even though future measurements will reveal additional information on $a$ and/or $b$. Specifically, the uncertainty free relations $p(a|m,b)$ require that any initial knowledge of the value of $m$ restricts the possible knowledge of $b$, so that all predictable probabilities will be positive and real. The uncertainty expressed by the probability distribution $|\langle b \mid m \rangle|^2$ is therefore a necessary consequence of the positivity of predicted probabilities and does not represent an arbitrary reduction of the correlations between $m$ and $b$. In quantum mechanics, the complex joint probabilities $\rho(a,b|m)$ of pure states $\mid m \rangle$ are therefore fully determined by fundamental laws of physics, despite the fact that the individual measurement outcomes of either $a$ or $b$ are unpredictable and random.

\section{Quantum determinism}

Weak measurement overcomes the uncertainty limit by combining initial information with final information. The operator formalism suggests that the weak values so obtained are indeed uncertainty free, but that the conditional probabilities for the different outcomes of projective weak measurements are complex. Since complex probabilities cannot be interpreted as relative frequencies, this result suggests that the realities of $(a,b)$ are fundamentally different from the realities of $(m,a)$ and from the realities of $(b,m)$. The meaning of complex joint probabilities must therefore be found elsewhere - specifically, in the relation between different possible measurements that can never be performed on the same system. Thus, the question is not a counter factual one (what would have happened in this specific system if we had measured $m$ instead of $b$?), but a statistical one (how to explain the statistics of $m$-measurements in terms of the statistics obtained in unrelated measurements of $a$ and $b$). Complex conditional probabilities are fundamental because they allow a reliable prediction of probabilities for measurement outcomes $m$ from the complex joint probabilities of the measurement outcomes $a$ and $b$ observed in separate measurements on the same kind of system \cite{Hof12a},
\begin{equation}
p(m) = \sum_{a,b} p(m|a,b) \rho(a,b|\psi). 
\end{equation}
Here, the conditional probability $p(m|a,b)$ is a universal relation between $a$, $b$, and $m$ that does not depend on the initial state $\psi$ or the measurements used to determine $\rho(a,b|\psi)$. No matter what the specific situation describe by the initial state $\psi$ is, these conditions can always be characterized by an experimentally observable complex joint probability of $a$ and $b$, and the relation between this joint probability and any other measurement $m$ is fundamentally defined by $p(m|a,b)$. Significantly, the complex conditional probability $p(m|a,b)$ is the {\it only} relation that quantum physics allows between $\hat{A}$, $\hat{B}$, and $\hat{M}$. The fundamental nature of this relation can be illustrated by expressing the operator $\hat{M}$ in terms of the weak value statistics for $a$ and $b$,
\begin{eqnarray}
\hat{M} &=& \sum_m M_m \mid m \rangle \langle m \mid 
\nonumber \\
&=& \sum_{a,b,m} M_m \; p(m|a,b) \mid b \rangle\langle b \mid a \rangle\langle a \mid.
\end{eqnarray}
The conditional probabilities $p(m|a,b)$ therefore describe the relation between the operator $\hat{M}$ and the operators $\hat{A}$ and $\hat{B}$ in terms of the spectral decompositions of the operators $\hat{A}$ and $\hat{B}$ given by their eigenstates $\{\mid a \rangle\}$ and $\{\mid b \rangle\}$. Based on this relation, the complete algebra of Hilbert space operators can be derived from the complex conditional probabilities. 

Since they are based on eigenstates and precise measurement results, complex conditional probabilities can provide a more detailed explanation of operator relations, where the non-classical properties often average out in the sum over all possible eigenstates and eigenvalues. In particular, the important analogies between the operator formalism and classical physics are mostly meaningful when the statistics is described in terms of expectation values. However, this is an approximation that is only valid when the effects of the necessary uncertainties are small enough to be neglected. It is therefore wrong to interpret the case where classical relations hold as absence of quantum correlations. Even in the most classical case, the microscopic relations between exact properties are given by complex probabilities of the form $p(m|a,b)$, which do not depend on the available information or the quantum state under consideration, and do not change into a more classical form as the uncertainties of the initial state increase. Therefore, it is always a mistake to assume that physical properties describe a separate material reality independent of interactions. Instead, even the classical limit should be understood as a reality of interactions and effects. It may be useful to remember how Dr. Johnson tried to refute Berkeley (and failed): the existence of a stone is known by kicking it, not by a dogmatic belief in the fundamental reality of its geometric shape in space. Interaction independent existence is a misinterpretation of reality, since reality is known by touch and sight. In the classical limit, the effects of interactions may be conveniently summarized by assigning approximate values to all properties, but it is important to keep in mind that the exact values of the properties $a$, $b$ and $m$ only refer to potential experiences of the object, which emerge gradually in the interaction with the object. In quantum physics, the effects of these interaction processes cannot be separated from reality, and the relations between different possible observations need to include the complex phases associated with the action of transformations. 

\section{Transformation dynamics}

Traditionally, much of physics is concerned with dynamics, and therefore with the changes of a specific property in time. However, it is important to recognize that quantum mechanics provides a description of time evolution that is actually incompatible with a description of dynamics as a continuous evolution of reality. The proper description of time evolution is given in terms of unitary operators that may be applied either to transform the states (Schroedinger picture) or the operators (Heisenberg picture). In fact, it is the transformation of operators that provides the correct interpretation of the dynamics: physical properties at different times are related to each other in a uniquely defined state independent way. The relation between observations of the same physical property at different times is therefore equivalent to the relation between observations of different physical properties at the same time. In this sense, Laplacian determinism is also valid in quantum mechanics, and the deterministic relation of $(a,b)$ and $m$ expressed by the complex conditional probability $p(m|a,b)$ is equally valid for the relation between initial position and momentum, and position at any later time. 

If $x_t$ is the position of a particle at time $t$ and $(x_0,p_0)$ is the position and the momentum at time zero, the motion of the particle is described by a complex conditional probability $p(x_t|x_0,p_0)$ that corresponds to the time dependent position operator $\hat{x}(t)$ of the Heisenberg picture,
\begin{equation}
\hat{x}(t) = \int x_t \; p(x_t|x_0,p_0) \mid p_0 \rangle\langle p_0 \mid x_0 \rangle \langle x_0 \mid \; dx_t.
\end{equation}
The complex conditional probability $p(x_t|x_0,p_0)$ shows that the time evolution of $x$ is still determined by the initial momentum, even though this dependence does not correspond to the linear evolution of classical physics. Instead, the complex phase of the probability changes with increasing rates as $x_t$ moves away from the classically predicted solution \cite{Hof12a}. If the position of the object is observed with limited resolution, the rapid oscillations of the complex probabilities average out, leaving only results close to the classical solution for $x_t$. Thus, the classical result emerges from complex conditional probabilities between the possible positions of the object, and not from a continuous movement of the center of mass along a mathematical line. Strictly speaking, the motion of objects observed in everyday life is therefore always associated with a complex probability distribution at the microscopic level, which is a necessary requirement of all motion observable at the macroscopic level. The familiar experiences of classical motion always involve interactions that allow us to observe the motion and therefore add sufficient noise to momentum and position to ensure that all observations can be described by the coarse grained versions of $p(x(t)|x_0,p_0)$. As shown in \cite{Hof12a}, such coarse graining quickly recovers the expectations of classical determinism, explaining how the classical description of motion emerges as a crude approximation of the quantum laws of motion expressed by the more precise complex conditional probabilities.

Importantly, complex probabilities make a constructive contribution to our understanding of quantum dynamics in the form of their imaginary parts. As discussed in \cite{Hof11b}, the complex phase of the conditional probability $p(m|a,b)$ describes the action required to optimally transform $a$ to $b$ along $m$, and this is the reason for the significance of the constant $\hbar$, which actually defines the ratio between the classical units of action and the phases of the complex conditional probabilities from which the action emerges in the approximate limit of classical physics. In the operator formalism, this relation appears in the well known Heisenberg equations of motion for the operators,
\begin{equation}
\frac{d}{dt} \hat{A} = - \frac{i}{\hbar} [\hat{A},\hat{H}].
\end{equation}
This equation is often discussed in terms of the correspondence between commutation relations and the Poisson brackets of classical physics. However, Eq.(\ref{eq:comrel}) shows that the commutation relations between $\hat{A}$ and $\hat{H}$ also represent an imaginary correlation between the observable $\hat{A}$ and the energy $\hat{H}$. It is therefore possible to find a macroscopic classical meaning in the imaginary correlations described by commutation relations. Specifically, the observable change of a property $\hat{A}$ in time is always equal to an imaginary correlation of this property and the energy, as given by the complex joint probability that characterizes the prior effects of the system,
\begin{eqnarray}
\label{eq:motion}
\frac{d}{dt} \langle \hat{A} \rangle &=& \frac{2}{\hbar} \mbox{Im}\left(\langle \hat{A}\hat{H} \rangle \right)
\nonumber \\ &=& \frac{2}{\hbar} \sum_{a,n} A_a E_n \; \mbox{Im}\left(\rho(n,a|\psi)\right).
\end{eqnarray}
Importantly, this relation only refers to the expectation value of $\hat{A}$, which can be interpreted as an estimate of the actual value $A_a$ based on the prior information $\psi$. The precision of the estimate is given by the uncertainty of $\hat{A}$ in $\psi$. In the classical limit, this precision is usually much higher than the precision of observations of $\hat{A}$, so the observed dynamics of $\hat{A}$ corresponds directly to the imaginary correlation. Eq.(\ref{eq:motion}) thus shows that classical motion invariably corresponds to microscopic imaginary correlations between the variable and the energy.

It is also possible to show that the equations of motion require imaginary correlations between observations of the same property at two different times. In particular, the velocity of an object of mass $m$ is given by $\hat{p}/m$, and the commutation relation of the momentum $\hat{p}$ and the position $\hat{x}$ is constant. Therefore, the two time correlation of position for an object freely moving through space has an imaginary part given by
\begin{equation}
\label{eq:twotime}
\mbox{Im}\left(\langle \hat{x}(t_2)\hat{x}(t_1) \rangle\right) = \frac{\hbar}{2 m} (t_2-t_1).
\end{equation}
Interestingly, this imaginary correlation increases in time, so that it would eventually become noticeable on a macroscopic level. However, quantum mechanics requires that an object can only move freely if there are no measurement interactions, and hence no measurements of the position between $t_1$ and $t_2$. If there is any interaction, it will reduce the correlations between the positions, just as it reduces the classical correlation between initial momentum and final position. It is therefore possible to derive the magnitude of the disturbances necessary for a continuous observation of motion from the evolution of imaginary correlations given by Eq. (\ref{eq:twotime}).

\section{Conclusions}

The results observed in weak measurements and the definition of measurement uncertainties introduced by Ozawa both indicate that the quantum formalism can provide an uncertainty free description of physical reality. Here, I argue that complex conditional probabilities are the adequate expression for the fundamental relations between the different physical properties of a system. In this context, a physical property should be understood as a possible effect of the system, so that the assignment of a specific value or outcome depends on the presence of an interaction that produces the corresponding effect. Significantly, the complex conditional probabilities observed in weak measurements provide a context independent description of the fundamental relation between  pairs of properties $(a,b)$ and any third property $m$, just as classical mechanics uni\-quely determines the relation between position and momentum at time zero and the position at any other time. Complex probabilities can thus provide a more detailed microscopic explanation of quantum mechanics, where the relation between different properties is fundamentally related to the action of transformations between the properties.

This work was supported by JSPS KAKENHI Grant Number 24540427.


\vfill

\begin{thebibliography}{10}


\bibitem{Joh07}
L. M. Johansen, Phys. Rev. A {\bf 76}, 012119 (2007).

\bibitem{Lun11}
J. S. Lundeen, B. Sutherland, A. Patel, C. Steward, and C. Bamber, Nature {\bf 474}, 188.

\bibitem{Lun12}
J. S. Lundeen and C. Bamber, Phys. Rev. Lett. {\bf 108}, 070402 (2012). 

\bibitem{Hof11b}
H.F. Hofmann, New J. Phys. {\bf 13}, 103009 (2011).

\bibitem{Hof12a}
H.F. Hofmann, New J. Phys. {\bf 14}, 043031 (2012).


\bibitem{Oza03}
M. Ozawa, Phys. Rev. A {\bf 67}, 042105 (2003).

\bibitem{Hal04}
M. J. W. Hall, Phys. Rev. A {\bf 69}, 052113 (2004).

\bibitem{Hos10}
A. Hosoya and Y. Shikano, J. Phys. A: Math. Gen. {\bf 43}, 025304 (2010).

\bibitem{Lun10}
A.P. Lund and H.M. Wiseman, New J. Phys. {\bf 12}, 093011 (2010).

\bibitem{Hof12b}
H.F. Hofmann, e-print arXiv:1205.0073 (2012).

\bibitem{Dir45}
P. A. M. Dirac, Rev. Mod. Phys. {\bf 17}, 195 (1945).



\end{thebibliography}
\end{document}